# Ground-to-UAV sub-Terahertz channel measurement and modeling

Da Li,[1] Peian Li,[1] Jiabiao Zhao,[1] Jianjian Liang,[2] Jiacheng Liu,[1] Guohao Liu,[1] Yuanshuai Lei,[1] Wenbo Liu,[1] Jianqin Deng,[3,*] Fuyong Liu,[4] and Jianjun Ma[1,*]

[1]School of Integrated Circuits and Electronics, Beijing Institute of Technology, Beijing 100081, China
[2]School of Automation, Beijing Institute of Technology, Beijing 100081, China
[3]Ceyear Technologies Co., Ltd., Qingdao 266555, China
[4]AutonCloud (Tianjin) Technology Co., Ltd., Tianjin 300384, China

**Abstract:** Unmanned Aerial Vehicle (UAV) assisted wireless communications have been expected to play a vital role in the next generation of wireless networks. UAVs can serve as either repeaters or data collectors within the communication link, thereby potentially augmenting the efficacy of communication systems. Despite their promise, the channel analysis and modeling specific to terahertz (THz) wireless channels leveraging UAVs remain under explored. This work delves into a ground-to-UAV channel at 140 GHz, with a specific focus on the influence of UAV hovering behavior on channel performance. Employing experimental measurements through an unmodulated channel setup and a geometry-based stochastic model (GBSM) that integrates three-dimensional positional coordinates and beamwidth, this work evaluates the impact of UAV dynamic movements and antenna orientation on channel performance. Our findings highlight the minimal impact of UAV orientation adjustments on channel performance and underscore the diminishing necessity for precise alignment between UAVs and ground stations as beamwidth increases.

## 1. Introduction

The future of wireless communication envisions a comprehensive network that seamlessly integrates space, air and ground elements, heralding an era of ubiquitous intelligence [1, 2]. At the forefront of this technological advancement is the integration of UAVs with wireless communication technologies. UAV-assisted THz and sub-terahertz (sub-THz) communication opens new avenues for enhancing channel propagation and user coverage, increasing the data transmission rates for payload links, and improving connectivity with the Internet of Things (IoT) [3]. This innovative approach is not only pivotal in augmenting the capabilities of existing 5G networks but also in shaping the landscape of future mobile communications, where UAVs emerge as versatile nodes - acting as mobile base stations, user equipments and repeaters, *etc*. [1, 4-6].

In this rapidly evolving landscape, UAVs have garnered immense interest for their versatility in both civilian and military spheres, underpinned by their affordability and adaptability. These aerial vehicles have been deployed across a spectrum of applications - from wildlife monitoring and border surveillance to facilitating wireless networks, executing search and rescue operations, managing delivery tasks, and conducting military maneuvers, among others [7-10]. Such widespread utilization underscores the importance of ground-to-UAV channel measurement and modeling, a cornerstone for advancing next-generation wireless communication technologies. This endeavor not only promises to revolutionize the way we deploy communication networks but also ensures that UAVs can perform optimally across the myriad roles they are tasked with.

Delving deeper into this critical research area, significant scholarly efforts have been noted, showcasing breakthroughs in ground-to-UAV channel measurement and modeling. For instance, comprehensive measurement campaigns have yielded data crucial for link design, with notable studies reporting on ground-to-UAV MIMO channels at various frequencies, highlighting the role of spatial diversity despite sparse multipath environments [11, 12]. Further, innovative models have been proposed, such as the non-stationary geometric sub-THz MIMO channel model and the stochastic path loss models for ultra-wideband spectrum, offering a nuanced understanding of air-to-ground propagation characteristics [13, 14]. Analysis of ground-to-UAV propagation at millimeter wave bands and in THz bands revealed distinct trends in channel sparsity with altitude and environmental influences, emphasizing the complex nature of these communications channels [15-17]. These insights, coupled with findings on the impact of urban and suburban settings on path loss and multipath component behavior, lay the groundwork for developing robust, high-performance UAV communication systems capable of operating under a wide range of conditions [18-29]. More complex factors are taken into account for THz wireless channel modeling, which incorporate environmental conditions, *i.e.*, temperature, humidity, and pressure, even the spatial jitter between the transmitting and receiving antennas [30]. Special scenarios, such as UAV-mounted absorbing metasurfaces channel modeling, attract researcher's attention, a comprehensive path loss model based on the Fresnel-Kirchhoff diffraction formula is developed for accurately calculating path loss in the specific scenario above [31], which is a typical application for UAV in next generation mobile communication. What's more, the complex relationship between antenna's directivity, power distribution and diverse antenna patterns is revealed [32], that is a significative reference for beam orientation in dynamic aerial communication links. As we continue to push the boundaries of UAV-assisted communication, these foundational studies illuminate the path toward seamless, efficient aerial networks, marking a significant stride in our journey toward truly ubiquitous wireless connectivity.

However, despite these strides, the intricacies of UAV-assisted channels, particularly in THz and sub-THz bands, demand continued exploration to bridge the gaps in our understanding and capabilities, especially in complex and dynamically changing environments. This work contributes to this endeavor by presenting a comprehensive study on the type of ground-to-UAV channels when the UAV is hovering, offering insights into the stability of channel transmission under such conditions. It compares the signal-to-noise ratio (SNR) across UAV trajectories and examines the impact of UAV movement on path loss, employing a novel channel model that incorporates dominant channel parameters, three-dimensional position coordinates, beamwidth and antenna orientation. Through measurement and modeling, this work aims to enhance our understanding of ground-to-UAV THz channel, paving the way for more reliable and efficient aerial communication systems in the era of ubiquitous connectivity.

The manuscript is structured as follows: Section II lays out the experimental setup of our investigation. In Section III, we delve into the analysis of power metrics, while Section IV is dedicated to the modeling of the channel and a thorough discussion, investigating the complexities of UAV-supported THz communications. Section V is focused on the examination of channel alignment. The manuscript concludes in Section VI, where we summarize our discoveries and propose avenues for future exploration.

## 2. Experimental setup

In the exploration of ground-to-UAV communication channels at the 140 GHz frequency band, our experimental setup was designed to capture the characteristics of sub-THz channels propagating in an aerial scenario. The main part of our measurement setup was an unmodulated channel configuration, as detailed in Fig. 1(a), consisting of a precisely engineered assembly of transmission and reception components.

For the ground station (transmitter) side, we utilized a Ceyear 1465D signal generator, renowned for its ability to produce signals up to 20 GHz. This signal was then frequency-multiplied using a Ceyear 82406B module (×12) to span the 110-170 GHz range. We focus on a 140 GHz continuous wave (CW) signal for our measurements, because it is one of the frequency ranges that has seen increasing interest for future wireless communication systems and the advances in semiconductor technologies (such as silicon-germanium (SiGe) BiCMOS and III-V compound semiconductors) [33, 34]. To transmit this signal, we employed an HD-1400SGAH25 horn antenna, complemented by a dielectric focal lens with a 10 cm focal length to enhance its gain to be 31 dBi (antenna gain of 25 dBi and an extra of 6 dB due to insertion of the lens), ensuring the directionality and strength of the signal were maintained across the transmission path.

On the UAV side, an identical horn antenna (gain of 25 dBi) was tasked with receiving the transmitted signal, without participation of focal lens. This antenna, in conjunction with a Ceyear 71718 power sensor connected to a Ceyear 2438PA power meter, formed the critical reception apparatus. To mitigate the influence of UAV body frame vibrations - induced by the interaction between the propellers and the UAV's main body [35] - on signal detection, the entire receiving assembly was insulated with multiple layers of flexible foam slices. The UAV's design, originating from the Beijing Institute of Technology (BIT), featured a 70 cm wheelbase and a weight of 4 kg, with 14-inch long propellers. This design allowed for the logging of 3D coordinates and pitch through a high-precision positioning module with two accelerometers/gyroscope sensors (ICM20699 and BMI055), one electronic compass (IST8310), which together provides a positional accuracy of ~ centimeter and an angular accuracy of 0.1 degree, enabling the monitor and inclusion of relative distance, attitude and orientation for our data analysis and channel modeling. What's more, the connection cable used to connect the UAV-mounted sensor head to the power meter is soft and lightweight (approximately 0.1 kg), does not significantly affect the flying stability of the UAV.

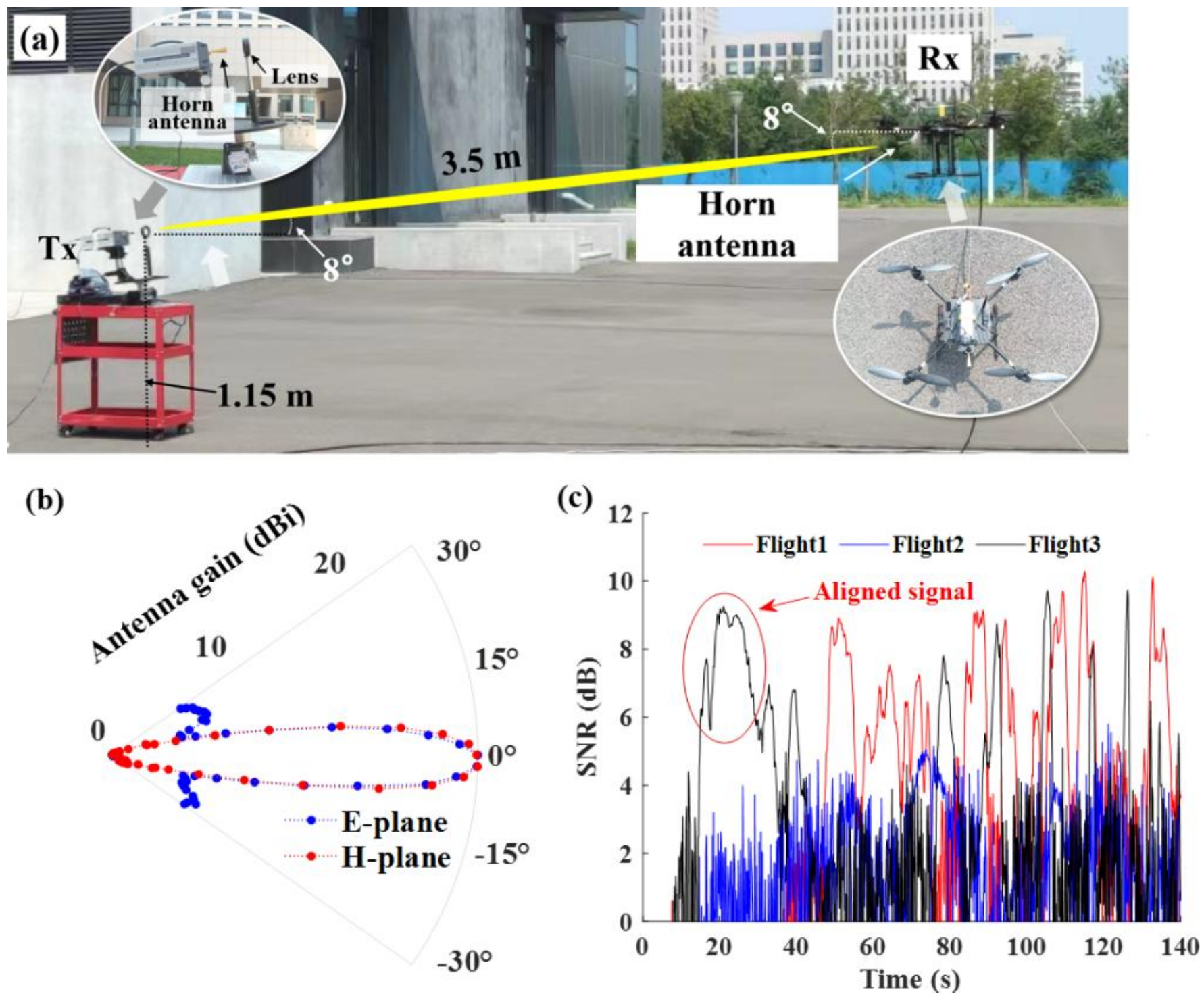

Fig. 1. (a) Picture of the setup for ground-to-UAV channel measurement; (b) Radiation pattern of the transmitter together with a horn antenna and a focal lens; (c) Evolution of detected signal-to-noise ratio (SNR).

For the purpose of maintaining optimal channel alignment, the transmitting antenna was set at an elevation angle of 8°, with the receiving antenna tilted at -8°. This configuration was critical to ensure that, as the UAV maneuvered through its flight path, the main lobe of the radiation pattern remained directed towards the receiver, as depicted in the radiation diagram shown in Fig. 1(b). The half-power beam-width (HPBW) of the antenna with a 10-cm lens is approximately 5.7°, measured over a 52 cm distance in a controlled laboratory environment, emphasizing the necessity for precise alignment to avoid significant reductions in received signal strength due to minor positional or orientational deviations [29].

Our experimental methodology included repeating the measurement process three times, corresponding to three UAV flights, to assess the stability and consistency of signal reception under varying conditions. Fig. 1(c) illustrates the signal-to-noise ratio (SNR) detected by our power meter during these flights. The maximum SNR, as in Fig. 1(c), is 10.28 dB. It is lower than the measured value in a static station (~11.1 dB), and also lower than the predicted value of 11.07 dB calculated using the Friis transmission formula. This indicates a near-perfect channel alignment we can achieve and a bit-error-ratio (BER) below $10^{-10}$ when amplitude shift keying (ASK) modulation scheme is employed [36]. The inherent instability associated with UAV hovering - attributed to limitations in the UAV's stabilization mechanisms - introduced a degree of variability in channel performance. This variability induces challenges in maintaining sustained optimal alignment and, by extension, consistent maximum power reception over time. This makes explorations of advanced stabilization techniques, adaptive beamforming algorithms, error-correction strategies necessary and required for mitigating the effects of UAV instability on channels. It should also be noted that the noise level, detected by our power meter, increases with UAV taking off. We think the vibration caused by the rotation of its blades with the rotor-blade modulation (RBM) effect may be responsible for that [35, 37, 38], that waiting to be explored continuously. Adaptive filtering, noise cancellation, and robust modulation schemes should be considered and employed to counteract the effects of the noise and ensure reliable data transmission in future.

## 3. Channel measurement and power analysis

The dynamics of ground-to-UAV communication, especially at the 140 GHz frequency band, necessitate a detailed analysis of channel power to comprehend the effects of UAV mobility on channel performance. We employ the cumulative distribution function (CDF) to dissect power measurements under two distinct operational states of the UAV: stationary, with the propellers deactivated, and hovering, wherein the UAV is subject to minor positional adjustments due to active propellers. The performance, observed in Fig. 2(a) and (b), is predicated on the premise of optimal channel alignment, as evidenced by the data captured within the red circle in Fig. 1(c).

It can be seen that the SNR in the stationary state adheres to a Rician distribution, indicative of a predominant line-of-sight (LOS) component's influence on the channel [39]. Conversely, the SNR characteristics of the hovering UAV channel are best described by a Weibull distribution (goodness-of-fit metric of 0.9641 for Weibull and 0.9325 for Rician), suggesting a departure from the strong, consistent LOS component observed in stationary conditions [40]. The Weibull distribution is renowned for its versatility in modeling a wide spectrum of signal fading phenomena, from mild to severe, making it particularly suited for capturing the nuanced channel performance variations in UAV communications. This variability under hovering conditions can be attributed to the severity and nature of fading, factors that are influenced more by the UAV's dynamic movements than by the static properties of the communication environment or the UAV's structural design [35]. The inherent instability of a hovering UAV, manifested through random movements in UAV location and variation of antenna orientation, underscores the potential for even minimal displacements or rotational adjustments to precipitate significant alterations in the signal path and, further, the SNR. Such observations underscore the limitations of traditional channel models, which typically presuppose a stationary state [11], to accurately encapsulate the complex dynamics of ground-to-UAV channel. Accordingly, we promote a paradigm shift in channel modeling for UAV channels, necessitating the incorporation of models that explicitly account for the UAV's motion and orientation shifts. This requirement for more sophisticated modeling techniques is not merely a reflection of the empirical data but a call for the development of analytical frameworks capable of accommodating the spatial variability introduced by UAV dynamics.

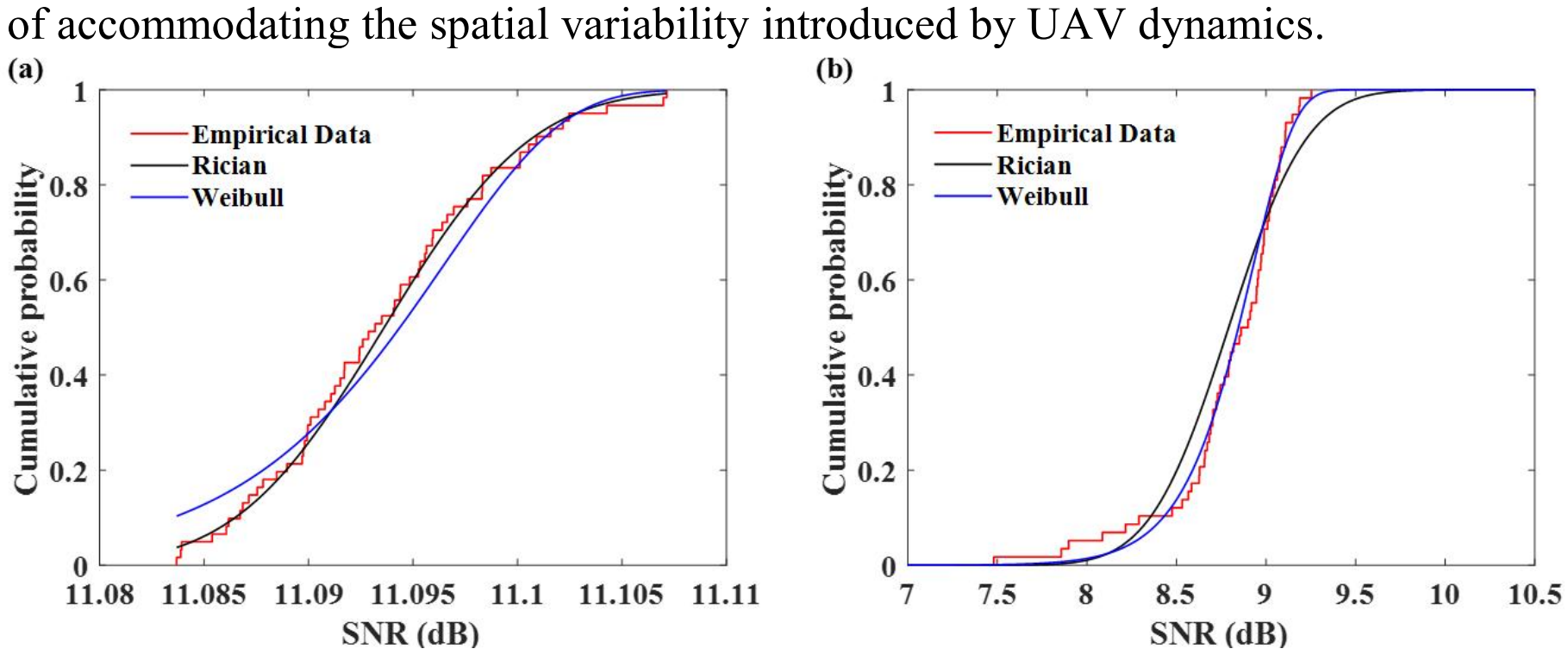

Fig. 2. CDF profile for received SNR with UAV in (a) stationary state and (b) hovering state.

**Table 1. Distribution parameters (in dB)**

| Parameters | Rician | Weibull |
|---|---|---|
| Scale parameter | 0.0056 | 8.93 |
| Rician factor K (Rician)/ shape parameter (Weibull) | 11.09 | 38.63 |

## 4. Channel modeling and evaluation

Our channel modeling endeavor aimed to investigate the characteristics of UAV dynamics and their impact on THz channel performance, with a particular emphasis on understanding how UAV hovering influences power variations within the wireless channel. By analyzing the power performance associated with the UAV's dynamic motion, we sought to quantify the effects of such movements on channel performance. This results, shown in Fig. 3(a) and (b), facilitated a granular examination of the UAV's three-dimensional (3D) motion, simplifying it into two-dimensional (2D) representations for clarity and ease of interpretation. Our focus was on scenarios where the transmitter and receiver were perfectly aligned, revealing that power variations remained within a 4 dB range (4.5 - 8.5 dB), indicative of a relatively stable wireless channel despite the inherent dynamic nature of UAV hovering.

To adequately model the ground-to-UAV channel, our approach embraced the complexities of three-dimensional UAV movement and orientation. We introduced a sophisticated three-dimensional geometry-based stochastic model (GBSM) dedicated to calculate path loss with consideration of these dynamic factors. This model incorporates the antenna radiation pattern, paying special attention to the beam alignment area during UAV flight - a critical region defined by a power reduction threshold of less than 3 dB. Our analysis, focused on the power variation insights from Fig. 1(c), highlighted a notable discrepancy in the beam alignment area's diameter (42 cm) compared to the HPBW (46 cm) at the UAV side, underscoring the tangible impact of UAV dynamic movement on antenna orientation and location, consequently, on channel propagation. The proposed model, as represented by the expression [41, 42],

$$SNR(dB) = P_T + G_{Tx} + G_{Rx} - 10\alpha\log_{10}(d) - \eta - N - \xi \quad (1)$$

which combines the interplay between various elements influencing channel transmission, including the path loss exponent [43-45] and small-scale fading [14, 24, 26] in hovering state as in Table 2. While, we do not consider the contribution of rain attenuation and atmospheric loss due to the sunny weather and short channel length during our measurement. Where, the units of $G_{Tx}$, $G_{Rx}$, $\eta$ and $\xi$ are all dB, the units of $P_T$ and $N$ are dBm, the unit of $d$ is meter, $\alpha$ is normalized. In this equation, the parameter $P_T$ represents the power launched out from transmitter, parameters $G_{Tx}$ and $G_{Rx}$ denote the antenna gains at the transmitter (together with the focal lens) and receiver sides, respectively, and parameter $d = \sqrt{a^2 + b^2 + (h_{UAV} - h_{Tx})^2}$ symbolizes the distance encapsulating both vertical (*a*) and horizontal (*b*) movement components, alongside the differential height between the UAV and the transmitter, with $h_{UAV}$ and $h_{Tx}$ as the height of UAV and transmitter respectively.

The model parameters - the noise power level $N$ (measured noise level by our power meter), spanning the path loss exponent $\alpha$, frequency-dependent path loss $\eta\,(dB) = 20\log_{10} f + 32.44$. The small-scale fading $\xi$ is characterized by a Weibull distribution—collectively offer a nuanced understanding of the path loss dynamics within the ground-to-UAV channel. Different from previous publications with $\xi$ following a Gaussian distribution [24], the probability density of $\xi$ in this work should be expressed by the function [49-51]

$$f_{\xi}(\omega;\varphi,\kappa) = \begin{cases} \frac{\kappa}{\varphi}(\frac{\omega}{\varphi})^{\kappa-1}e^{-(\omega/\varphi)^{\kappa}}, & w \geq 0 \\ 0, & w < 0 \end{cases} \quad (2)$$

due to its following on Weibull distribution in hovering state. The parameter $\omega$ is an independent variable, $\varphi$ and $\kappa$ are the scale and shape parameters of Weibull distribution, respectively, which must be greater than zero. The Weibull distribution, with $\omega$ as the independent variable, serves as a versatile tool for modeling the variability in signal fading. The scale parameter, $\varphi$, correlates with the average power of the fading signal, whereas the shape parameter, $\kappa$ reflects the severity of small-scale fading. Intriguingly, as the severity of fading intensifies, the value of $\kappa$ diminishes, offering a window into the distribution's adaptability. This adaptability is further exemplified by the distribution's transformational ability, where specific values of $\kappa$ (notably, 1.0 for an exponential distribution and 2.0 for a Rayleigh distribution) highlight its flexible application across different fading scenarios.

**Table 2. Theoretical models for UAV channel**

| Theoretical model | Ref. |
|---|---|
| $PL(dB) = 10\alpha\log_{10}(d)$, with $\alpha = 1.922, 1.5$, or $2.05$ | [43-45] |
| $P_R(dBm) = P_t + G_{UAV_1} + G_{UAV_2} + 10\log_{10}(\frac{\lambda}{4\pi d})^{\alpha}$, with $\alpha = 2.6$ | [46] |
| $PL(dB) = PL(d_0) + 10\alpha\log_{10}(\frac{d}{d_0})$, with $d_0$ as the reference distance | [22] |
| $P_R(dBm) = P_T + G_{Tx} + G_{Rx} - L_F - L_R - L_A - L_O$, with $L_F$ as free space path loss, $L_R$ as rain attenuation, $L_A$ as gaseous atmospheric loss, and $L_O$ as fading losses. | [47] |
| $PL(dB) = PL(d_0) + 10\alpha\left(\log_{10}\frac{d}{d_0}\right) - \log_{10}\frac{\Delta h}{h_{opt}} + C_p + 10x\log_{10}(\frac{f_c + \Delta f}{f_c}) + \xi$ | [14] |
| $PL(dB) = \alpha(h_{UAV})10\log_{10}(d) + \beta(h_{UAV}) + \xi$ | [24] |
| $PL(dB) = \alpha 10\log_{10}(d) + A(\phi - \phi_0)\exp\left(-\frac{\phi - \phi_0}{B}\right) + \eta_0 + \xi$ | [26] |
| $PL(dB) = PL(d_0) + 10n\log_{10}\left(\frac{d}{d_0}\right) + X_{\sigma}(d)$ | [48] |

In our study, we primarily focused on the UAV in a hovering state where the stability of the UAV is controlled within tight bounds. Specifically, the rolling angle of the UAV is maintained at approximately 2.36 degree during operations. The de-polarization loss can be quantified as $L_{dp}(dB) = -20\log_{10}(\cos 2.36) = 0.0074\ dB$, which is negligible and indicates the influence of antenna de-polarization is not required in our theoretical model as Eq. (1).

We keep the parameter setting of $P_T$ as 0 dBm, $\alpha$ as 1.5 [44], $N$ as -38 dBm, $\eta$ as 75.3 dB, $\kappa$ and $\varphi$ of $\xi$ being 8.93 and 38.63, respectively. The $G_{Tx}$ changes from 31 dBi (antenna gain of 25 dBi and an extra of 6 dB due to insertion of the focal lens) when UAV moves inside the beam path and $G_{Rx}$ changes from 25 dBi due to the variation of antenna orientation caused by UAV dynamics. The calculation of the antenna gains are based on the expression $G = G_0 \exp\left[-(\frac{\Phi-\Phi_0}{\sigma_\Phi})^2 - (\frac{\theta-\theta_0}{\sigma_\theta})^2\right]$, where the maximum gain can be obtained by formula $G_0 = 10\log_{10}(\beta\frac{4\pi A_p}{\lambda^2})$, with $\beta$ as the antenna efficiency and $A_p$ as the physical aperture area of the antenna [32]. The parameters $\Phi_0$ and $\theta_0$ denote the azimuth and elevation angles when a perfect on-axis alignment condition achieved, while $\Phi$ and $\theta$ represent the azimuth and elevation angles when the UAV is in a hovering state. The parameters $\sigma_\Phi$ and $\sigma_\theta$ describe the beamwidth in the azimuth and elevation directions, respectively. For the Tx side with a lens included, the beamwidth is 5.7°, and for the Rx side, it is 12°.

The comparison, between measured data (Fig. 3(a-b)) and theoretical results as illustrated in Fig. 3(c-d), underscores a difference up to 8 dB. Such a large discrepancy undermines confidence in the model's predictions, making it unreliable for designing and optimizing UAV communication systems. This discrepancy also casts doubt on the initial applicability of our proposed model under the tested conditions. Thus, a pressing need for model optimization to more accurately reflect the complex interplay between UAV and the channel is required definitely.

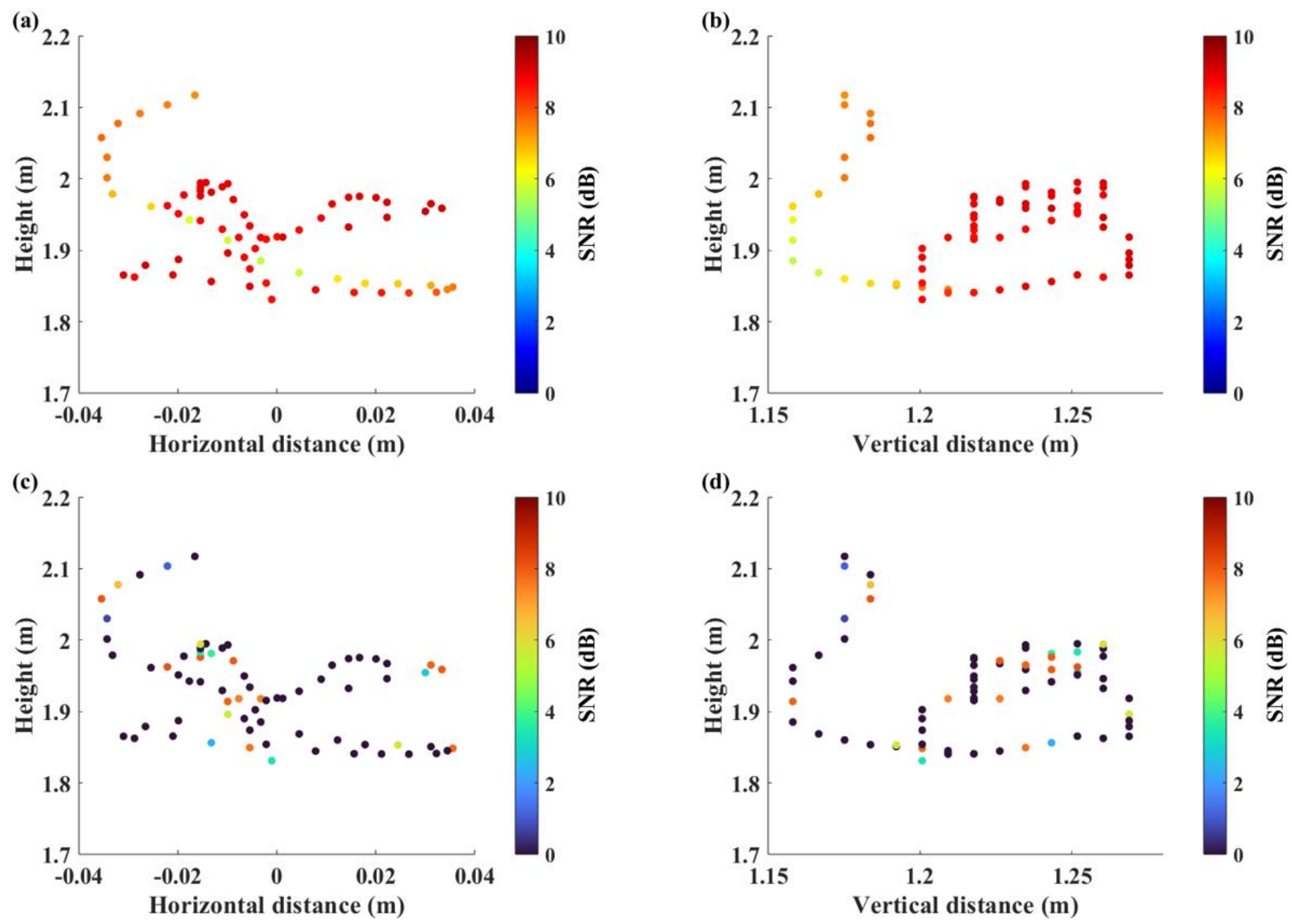

Fig. 3. (a-b) Measured and (c-d) calculated SNR with respect to 2D UAV trajectory.

To enhance the predictive accuracy of our model, we engaged in a detailed statistical analysis using Statistical Product and Service Solutions (SPSS) [52] to investigate the relationship between UAV movement, antenna orientation, and channel performance. The correlation analysis in SPSS is performed using the Pearson correlation coefficient formula (performed in linear scale), which is given by the formula

$$\rho_{X,Y} = \frac{\mathrm{Cov}(X,Y)}{\sigma_X \sigma_Y} \tag{3}$$

where, $\mathrm{Cov}(X,Y)$ is covariance, $\sigma_X$ and $\sigma_Y$ are the variance of variables $X$ and $Y$ respectively. The SPSS evaluates the linear relationship between both variables. A negative correlation coefficient indicates an inverse relationship, which is crucial for understanding opposing trends in our data, and can guide the development of more accurate channel models. The application of this statistical analysis is invaluable for identifying and quantifying relationships between complex variables, which helps for establishing robust channel models by concentrating on significant factors [53-55].

The analysis, as outlined in Table 3, revealed a notably low correlation between power and Rx's antenna orientation, that indicates the power performance of the ground-to-UAV channel is slightly affected by minor adjustments of antenna orientation of the Rx due to the movement of the UAV. Such findings highlight the antenna system's resilience, capable of maintaining consistent performance despite the range of orientation changes typically experienced during UAV hovering, affirming the system's adaptability and stability under dynamic flight conditions.

**Table 3. SPSS correlation analysis (in linear scale)**

| | Horizontal movement | Vertical movement | Height | Antenna orientation | Power |
|---|---|---|---|---|---|
| Horizontal movement | 1 | -0.836** | 0.032 | -0.021 | -0.641** |
| Vertical movement | | 1 | 0.443** | -0.018 | 0.798** |
| Height | | | 1 | -0.118 | 0.289** |
| Antenna orientation | | | | 1 | 0.061 |
| Power | | | | | 1 |

**. The correlation is significant at level 0.01 (2-tailed)

Based on the above correlation analysis, we proceeded to calibrate our theoretical model by discounting the previously overestimated impact of antenna orientation induced by UAV hovering. The calibrated model is under the condition that the $G_{Tx}$ still changes from 31 dBi based on the variation of the azimuth and elevation angles, while the $G_{Rx}$ always keeps at 25 dBi for a fixed antenna orientation. The results in Fig. 4(a) and (b), exhibits a much closer alignment with actual measurements, with discrepancies now reduced to be less than 2 dB as shown in Fig. 4(c-d). This improved accuracy, especially evident in the beam alignment area, reaffirms the limited correlation between detected power and antenna orientation under our specific experimental conditions. Consequently, this calibration not only enhances the accuracy of our channel model but also offers practical implications for the design and operational strategies of UAV communication systems, suggesting that relatively minor adjustments for antenna orientation may not necessitate the level of attention previously assumed.

To make the illustration clearer, we show the mean and root-mean-square (RMS) values of the difference between measured and calculated results in Table 4. It is clear that the difference is reduced from the point of mean and RMS values, when the theoretical model (Eq. (1)) is employed without (wo) antenna orientation considered. Further, we also want to check the impact of small-scale fading ($\xi$) in Eq. (1) on the calculation result. As shown in Table 4, the presence of this small-scale fading leads to a small reduction of the calculation accuracy in the mean value. So we employ the Eq. (1) with the small-scale fading ($\xi$) for future theoretical calculations in the next section.

To further clarify the results, we present the mean and root-mean-square (RMS) values of the differences between measured and calculated results in Table 4. It is evident that these differences are reduced based on the mean and RMS values when the theoretical model (Eq. (1)) is used without considering antenna orientation. Additionally, we examined the impact of small-scale fading ($\xi$) in Eq. (1) on the calculation results. As shown in Table 4, the inclusion of small-scale fading leads to a slight increasing in calculation accuracy. Consequently, the Eq. (1) with small-scale fading should be considered for future theoretical calculations.

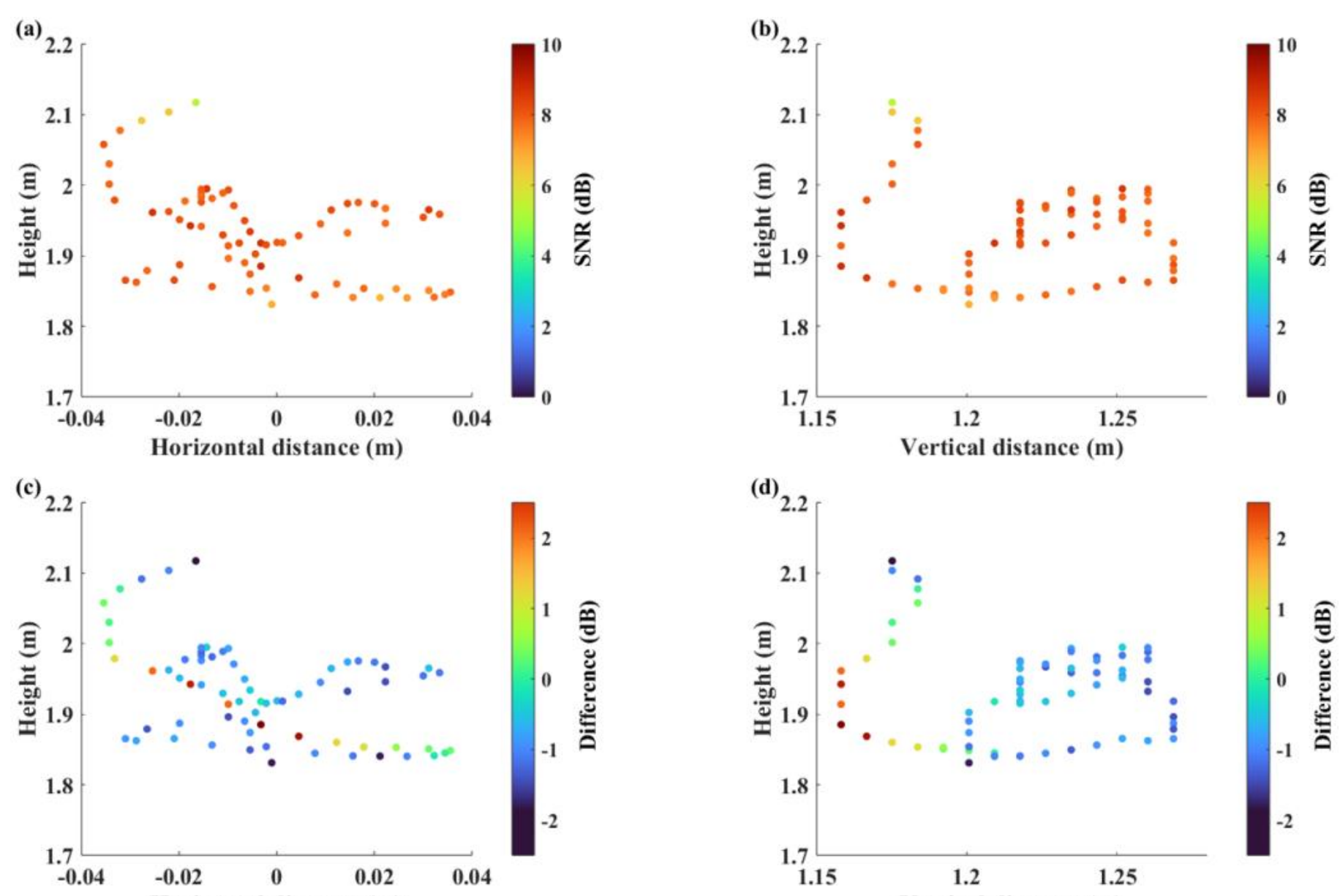

Fig. 4. (a-b) Predicted power level with respect to 2D UAV trajectory; (c-d) Difference between measured data and predicted results.

**Table 4. Mean and RMS values of the difference between experimental and calculation results (w: with, wo: without)**

| Theoretical model of Eq. (1) | | Mean (dB) | RMS (dB) |
|---|---|---|---|
| Small-scale fading | Antenna orientation | | |
| w | w | -6.541 | 17.208 |
| w | wo | -0.482 | 1.482 |
| wo | w | -6.566 | 17.218 |
| wo | wo | -0.526 | 0.977 |

In addition, we conduct comparative analyses against other theoretical models as depicted in the references [22, 48], to further validate the robustness of our proposed channel model. Fig. 5 and Table 5 present the power levels predicted by these models, juxtaposed with our measurement data. The results distinctly reveal that the discrepancies between the measured and predicted power levels are larger for the models proposed in [22, 48], as compared to the differences observed with our model. This notable reduction in deviation further underscores the superior accuracy and alignment of our model with the experimental measurements.

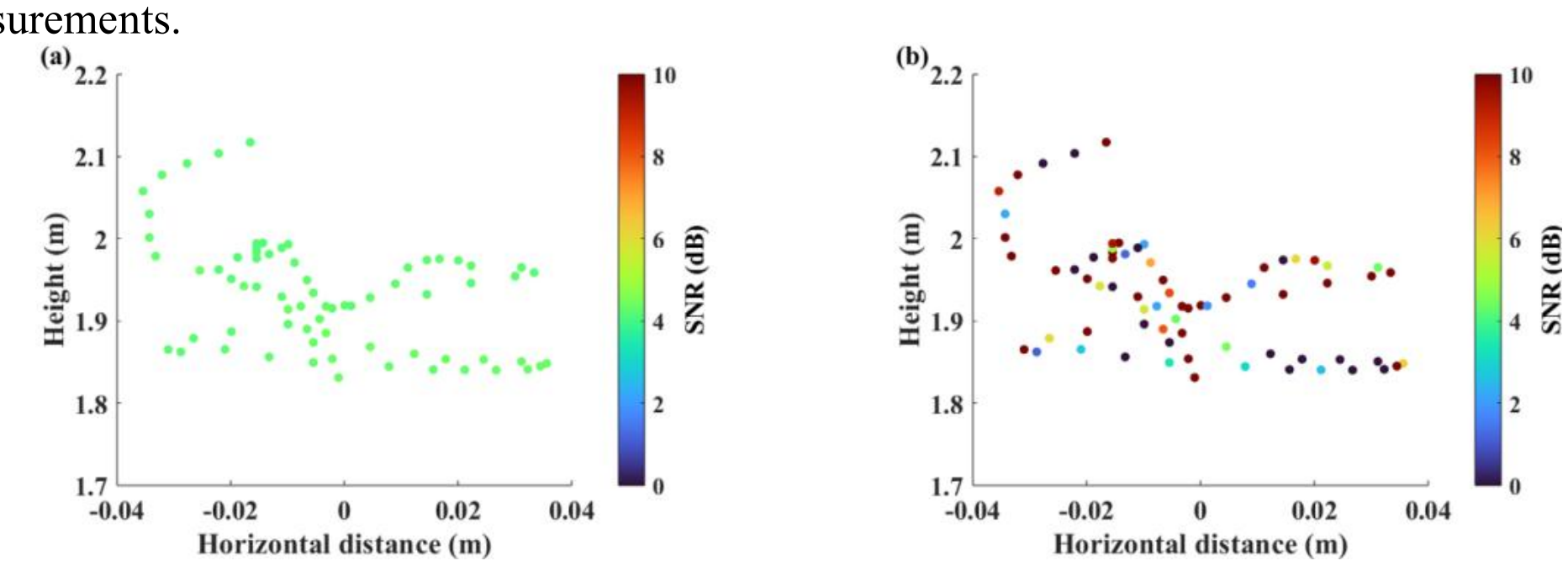

Fig. 5. Predicted power level with respect to 2D UAV trajectory: (a) by [22]; (b) by [48].

**Table 5. Mean and RMS values of the difference between experimental and calculation results by theoretical models**

| Theoretical models | Mean (dB) | RMS (dB) |
|---|---|---|
| Reference [22] | -4.071 | 12.454 |
| Reference [48] | -1.841 | 19.938 |

## 5. Channel alignment evaluation

The future evolution from ground-based mobile Ad Hoc Networks to aerial counterparts underscores the growing emphasis on UAV performance optimization and the expansion of operational coverage areas [56]. However, the inherently dynamic nature of UAV movement introduces significant challenges in maintaining precise beam alignment - a critical factor for sustaining robust communication links within these networks. The agility and mobility of UAVs necessitate sophisticated strategies to achieve and maintain alignment between the UAV (as the receiver) and the ground station (as the transmitter), especially when considering the need to counteract considerable path loss over vast distances [57, 58]. This challenge is not merely technical but foundational to the design and implementation of effective UAV networks, where the precision in beam positioning and alignment underpins the overall reliability and efficiency of communication.

In addressing the intricacies of UAV-induced dynamics on communication efficacy, the concept of a 3-dB range is defined as a crucial analytical parameter. This metric serves to delineate a beam-width boundary within which the attenuation of channel power is restricted to a maximum of 3 dB, even in the face of UAV motion. The exploration of the 3-dB range across varying distances and associated HPBW illuminates the intricate balance between beamwidth and channel alignment tolerance. By applying path loss models as in Eq. (1) to ascertain the diameter of the beam-aligned region (3-dB range), we gain insights into the impact of UAV movement on the stability and efficiency of the communication link. The result is shown in Fig. 6, where there is a linear correlation between the 3-dB range and the HPBW of the channel beam across a spectrum of distances, from 10 m to 200 m. This linear relationship provides a pivotal understanding: as the HPBW widens, the exigency for meticulous alignment between the UAV and the ground station diminishes. This relaxation in alignment precision underscores the receiver's (UAV's) capacity to receive the transmitted signal, even in instances where it is not perfectly situated within the transmitter's main lobe. Such a characteristic fundamentally enhances the resilience of the ground-to-UAV channel against minor positional or angular discrepancies, thereby streamlining the operational protocols necessary to uphold effective communication links.

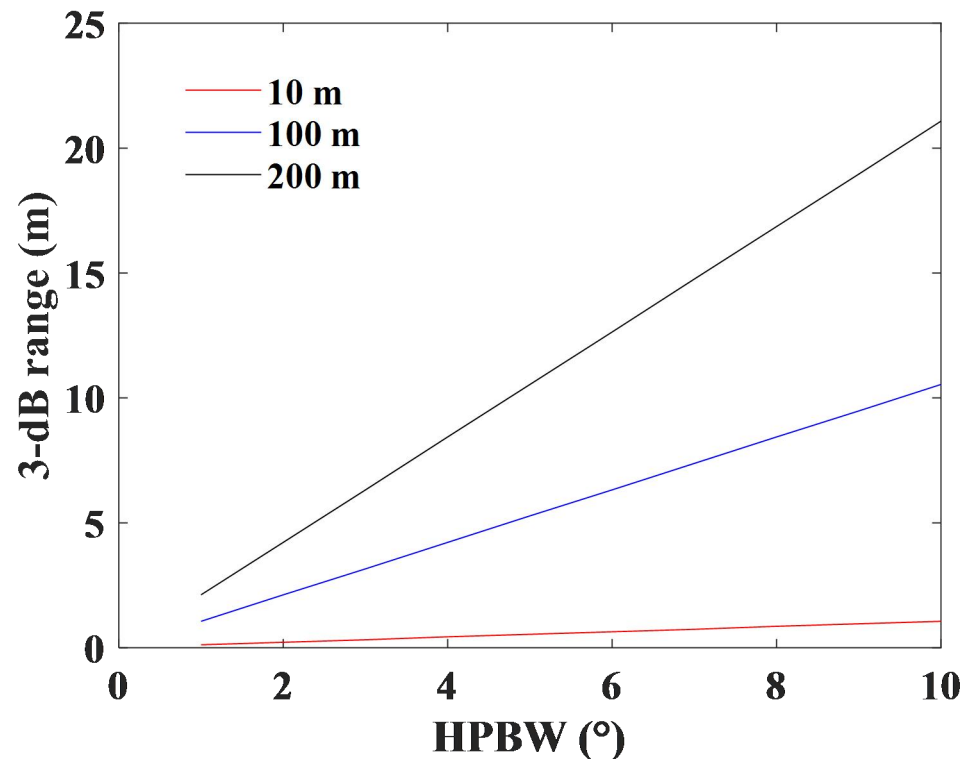

Fig. 6. Variation of 3-dB range with respect to HPBW over different channel distances.

## 6. Conclusion

In this work, we delved into the evolving landscape of wireless communication systems, underscored by the integration of UAVs and THz communication technologies. We conducted measurement on a ground-to-UAV THz channel at BIT and highlighted the effects of UAV dynamic movement and antenna orientation on channel performance. We find that the received channel power adheres to a Weibull distribution. By employing a novel GBSM, we were able to quantify the effects of UAV hovering on path loss and analyze the signal-to-noise ratio (SNR) across various UAV trajectories. Our findings indicate a marginal impact of UAV orientation on channel performance, which is instrumental in the development of more resilient and effective UAV communication systems. Additionally, the channel alignment analysis elucidates the relationship between beamwidth and channel alignment tolerance amidst UAV movement, which reveals that precise alignment becomes less critical as the HPBW increases, offering valuable insights into the design and operational strategies of UAV-assisted communication networks.

We think these findings have practical implications for the design and operation of UAV communication systems. Especially, understanding that minor orientation changes have a negligible impact on signal strength simplifies the system's requirements, potentially reducing the need for complex, real-time orientation adjustments or sophisticated antenna designs aimed at countering such changes. This allows designers and operators to focus more on other critical aspects, such as environmental factors, UAV altitude, and larger-scale movements, which might have a more pronounced effect on communication quality and reliability.

**References**

1. Z. Zhang, Y. Xiao, Z. Ma, et al., "6G Wireless Networks: Vision, Requirements, Architecture, and Key Technologies," IEEE Veh. Technol. Mag. 14(3), 28-41 (2019).
2. B. Rong, "6G: The Next Horizon: From Connected People and Things to Connected Intelligence," IEEE Wireless Commun. 28(5), 8-8 (2021).
3. X. You, C.-X. Wang, J. Huang, *et al.*, "Towards 6G wireless communication networks: vision, enabling technologies, and new paradigm shifts," Sci. China Inform. Sci. 64(1), 1-74 (2021).
4. H. Chang, C. X. Wang, Y. Liu, *et al.*, "A Novel Nonstationary 6G UAV-to-Ground Wireless Channel Model With 3-D Arbitrary Trajectory Changes," IEEE Internet Things J. 8(12), 9865-9877 (2021).
5. S. Hayat, E. Yanmaz and R. Muzaffar, "Survey on Unmanned Aerial Vehicle Networks for Civil Applications: A Communications Viewpoint," IEEE Commun. Surv. Tutorials 18(4), 2624-2661 (2016).
6. R. M. Rolly, P. Malarvezhi and T. D. Lagkas, "Unmanned aerial vehicles: Applications, techniques, and challenges as aerial base stations," Int. J. Distrib. Sens. Netw. 18(9), 1-24 (2022).
7. A. A. Khuwaja, Y. Chen, N. Zhao, *et al.*, "A Survey of Channel Modeling for UAV Communications," IEEE Commun. Surv. Tutorials 20(4), 2804-2821 (2018).
8. X. Cai, A. Gonzalez-Plaza, D. Alonso, *et al.*, "Low altitude UAV propagation channel modelling," in *2017 11th European Conference on Antennas and Propagation (EUCAP)* (2017), pp. 1443-1447.
9. S. A. H. Mohsan, N. Q. H. Othman, Y. Li, *et al.*, "Unmanned aerial vehicles (UAVs): practical aspects, applications, open challenges, security issues, and future trends," Intel. Serv. Robot. 109-137 (2023, doi: https://doi.org/10.1007/s11370-022-00452-4).
10. U. C. Cabuk, M. Tosun, O. Dagdeviren, *et al.*, "Modeling Energy Consumption of Small Drones for Swarm Missions," IEEE Trans. Intell. Transp. Syst. (doi: 10.1109/TITS.2024.3350042).
11. N. Stofik, D. W. Matolak and A. Sahin, "Measurement and Modeling of Low-Altitude Air-Ground Channels in Two Frequency Bands," in *2022 Integrated Communication, Navigation and Surveillance Conference (ICNS)* (2022), pp. 1-10.

12. T. J. Willink, C. C. Squires, G. W. K. Colman, *et al.*, “Measurement and Characterization of Low-Altitude Air-to-Ground MIMO Channels,” IEEE Trans. Veh. Technol. 65(4), 2637-2648 (2016).
13. K. Zhang, H. Wang, C. Zhang, *et al.*, “Three-Dimensional Non-Stationary Geometry-Based Modeling of Sub-THz MIMO Channels for UAV Air-to-Ground Communications,” in *ICC 2023 - IEEE International Conference on Communications* (2023), pp. 2069-2074.
14. W. Khawaja, I. Guvenc and D. Matolak, “UWB Channel Sounding and Modeling for UAV Air-to-Ground Propagation Channels,” in *2016 IEEE Global Communications Conference (GLOBECOM)* (2016), pp. 1-7.
15. W. Khawaja, O. Ozdemir and I. Guvenc, “UAV Air-to-Ground Channel Characterization for mmWave Systems,” in *2017 IEEE 86th Vehicular Technology Conference (VTC-Fall)* (2017), pp. 1-5.
16. P. S. Bithas, V. Nikolaidis, A. G. Kanatas, *et al.*, “UAV-to-Ground Communications: Channel Modeling and UAV Selection,” IEEE Trans. Commun. 68(8), 5135-5144 (2020).
17. Y. Li, N. Li and C. Han, “Ray-tracing Simulation and Hybrid Channel Modeling for Low-Terahertz UAV Communications,” in *ICC 2021 - IEEE International Conference on Communications* (2021), pp. 1-6.
18. M. Simunek, P. Pechac and F. Perez Fontan, “Excess Loss Model for Low Elevation Links in Urban Areas for UAVs,” Radioengineering 20(3), 561-568 (2011).
19. E. W. Frew and T. X. Brown, “Airborne Communication Networks for Small Unmanned Aircraft Systems,” Proc. IEEE 96(12), 2008-2027 (2008).
20. E. Yanmaz, S. Hayat, J. Scherer, *et al.*, “Experimental performance analysis of two-hop aerial 802.11 networks,” in *2014 IEEE Wireless Communications and Networking Conference (WCNC)* (2014), pp. 3118-3123.
21. E. Yanmaz, R. Kuschnig and C. Bettstetter, “Channel measurements over 802.11a-based UAV-to-ground links,” in *2011 IEEE GLOBECOM Workshops (GC Wkshps)* (2011), pp. 1280-1284.
22. E. Yanmaz, R. Kuschnig and C. Bettstetter, “Achieving air-ground communications in 802.11 networks with three-dimensional aerial mobility,” in *2013 Proceedings IEEE INFOCOM* (2013), pp. 120-124.
23. T. Tavares, P. Sebastião, N. Souto, *et al.*, “Generalized LUI Propagation Model for UAVs Communications Using Terrestrial Cellular Networks,” in *2015 IEEE 82nd Vehicular Technology Conference (VTC2015-Fall)* (2015), pp. 1-6.
24. R. Amorim, H. Nguyen, P. Mogensen, *et al.*, “Radio Channel Modeling for UAV Communication Over Cellular Networks,” IEEE Wireless Commun. Lett. 6(4), 514-517 (2017).
25. S. Dang, O. Amin, B. Shihada, *et al.*, “What should 6G be?,” Nat. Electron. 3(1), 20-29 (2020).
26. A. Al-Hourani and K. Gomez, “Modeling Cellular-to-UAV Path-Loss for Suburban Environments,” IEEE Wireless Commun. Lett. 7(1), 82-85 (2018).
27. M. A. Karabulut, “Study of Power and Trajectory Optimization in UAV Systems Regarding THz Band Communications with Different Fading Channels,” Drones 7(8), 1-17 (2023).
28. H. S. Cankurtaran, A. Kachroo, W. Choi, *et al.*, “Propeller Effects on mmWave UAV Channels: A Statistical and Empirical Modeling Study,” in *2022 IEEE International Black Sea Conference on Communications and Networking (BlackSeaCom)* (2022), pp. 30-35.
29. P. Wang, J. Fang, W. Zhang, *et al.*, “Fast Beam Training and Alignment for IRS-Assisted Millimeter Wave/Terahertz Systems,” IEEE Trans. Wireless Commun. 21(4), 2710-2724 (2022).
30. A.-A. A. Boulogeorgos, E. N. Papasotiriou, and A. Alexiou, “Analytical Performance Evaluation of THz Wireless Fiber Extenders,” in *2019 IEEE 30th Annual International Symposium on Personal, Indoor and Mobile Radio Communications (PIMRC)* (2019), pp. 1-6.
31. A. Pitilakis, D. Tyrovolas, P. V. Mekikis, *et al.*, “On the Mobility Effect in UAV-Mounted Absorbing Metasurfaces: A Theoretical and Experimental Study,” IEEE Access 11, 79777-79792 (2023).
32. C. A. Balanis, *Antenna Theory: Analysis and Design* (Wiley, 2015).
33. Y. Xing and T. S. Rappaport, “Propagation Measurement System and Approach at 140 GHz-Moving to 6G and Above 100 GHz,” in *2018 IEEE Global Communications Conference (GLOBECOM)* (2018), pp. 1-6.
34. K. Sengupta, T. Nagatsuma and D. Mittleman, “Terahertz integrated electronic and hybrid electronic–photonic systems,” Nat. Electron. 1(12), 622–635 (2018).
35. H. S. Cankurtaran, A. Kachroo, W. Choi, *et al.*, “Propeller Effects on mmWave UAV Channels: A Statistical and Empirical Modeling Study,” in *Secondary Propeller Effects on mmWave UAV Channels: A Statistical and Empirical Modeling Study* (2022), pp. 30-35.
36. J. Ma, R. Shrestha, L. Moeller, *et al.*, “Invited Article: Channel performance for indoor and outdoor terahertz wireless links,” APL Photonics 3(5), 1-12 (2018).
37. F. T. Faul, D. Korthauer and T. F. Eibert, “Impact of Rotor Blade Rotation of UAVs on Electromagnetic Field Measurements,” IEEE Trans. Instrum. Meas. 70(1), 1-9 (2021).
38. J. Bredemeyer and T. Schrader, “Employing UAS to perform low altitude navaids measurements”, in *International Flight Inspection Symposium (IFIS)* (2018), pp. 1-17.
39. P. Taiwo and A. Cole-Rhodes, “Performance of a Blind Adaptive Digital Beamformer on a Multi-user Rician Channel,” in *2021 55th Annual Conference on Information Sciences and Systems (CISS)* (2021), pp. 1-6.
40. J. A. Gay-Fernández and I. Cuiñas, “Using Weibull distribution to modeling short-term variations in decidious-forested radio channels,” in *2014 IEEE Antennas and Propagation Society International Symposium (APSURSI)* (2014), pp. 1568-1569.
41. H. Kang, J. Joung, and J. Kang, “Beamwidth of Base Stations for Maximizing Coverage of Aerial Users,” in *2019 Eleventh International Conference on Ubiquitous and Future Networks (ICUFN)* (2019), pp. 108-110.
42. H. He, S. Zhang, Y. Zeng, *et al.*, “Joint Altitude and Beamwidth Optimization for UAV-Enabled Multiuser Communications,” IEEE Commun. Lett. 22(2), 344-347 (2018).
43. J. Allred, A. B. Hasan, S. Panichsakul, *et al.*, “SensorFlock: An Airborne Wireless Sensor Network of Micro-Air Vehicle,” in *the 5th international conference on Embedded networked sensor systems* (2007), pp. 117–129.
44. A. Shaw and K. Mohseni, “A Fluid Dynamic Based Coordination of a Wireless Sensor Network of Unmanned Aerial Vehicles: 3-D Simulation and Wireless Communication Characterization,” IEEE Sens. J. 11(3), 722-736 (2011).
45. N. Ahmed, S. S. Kanhere and S. Jha, “On the importance of link characterization for aerial wireless sensor networks,” IEEE Commun. Mag. 54(5), 52-57 (2016).
46. N. Goddemeier and C. Wietfeld, “Investigation of Air-to-Air Channel Characteristics and a UAV Specific Extension to the Rice Model,” in *2015 IEEE Globecom Workshops (GC Wkshps)* (2015), pp. 1-5.
47. C. Yan, L. Fu, J. Zhang, *et al.*, “A Comprehensive Survey on UAV Communication Channel Modeling,” IEEE Access 7, 107769-107792 (2019).
48. A. Kachroo, Collin A. Thornton, Md. Arifur Rahman Sarker, *et al.*, “Emulating UAV Motion by Utilizing Robotic Arm for mmWave Wireless Channel Characterization,” IEEE Trans. Antennas Propag. 69(10), 6691-6701 (2021).
49. M. You, H. Sun, J. Jiang, *et al.*, “Effective Rate Analysis in Weibull Fading Channels,” IEEE Wireless Commun. Lett. 5(4), 340-343 (2016).

50. M. Gupta, J. Anandpushparaj, P. Muthuchidambaranathan, *et al.*, "Outage Performance Comparison of Dual-Hop Half/Full Duplex Wireless UAV System over Weibull Fading Channel," in *2020 International Conference on Wireless Communications Signal Processing and Networking (WiSPNET)* (2020), pp. 177-181.
51. N. C. Sagias and G. S. Tombras, "On the cascaded Weibull fading channel model," J. Franklin Inst. 344(1), 1-11 (2007).
52. E. Raeva, V. Mihova and I. Nikolaev, "Using SPSS for Solving Engineering Problems," in *2019 29th Annual Conference of the European Association for Education in Electrical and Information Engineering (EAEEIE)* (2019), pp. 1-6.
53. Stijn Sackesyn, Chonghuai Ma, Joni Dambre, *et al.*, "Experimental realization of integrated photonic reservoir computing for nonlinear fiber distortion compensation," Opt. Express 29(20), 30991-30997 (2021).
54. Jessica C. Ramella-Roman, Scott A. Prahl, and Steve L. Jacques, "Three Monte Carlo programs of polarized light transport into scattering media: part I," Opt. Express 13(12), 4420-4438 (2005).
55. Courtenay LA, González-Aguilera D, Lagüela S, *et al.*, "Hyperspectral imaging and robust statistics in non-melanoma skin cancer análisis," Biomed. Opt. Express 12(8), 5107-5127 (2021).
56. J. Wang, C. Jiang, Z. Han, *et al.*, "Taking Drones to the Next Level: Cooperative Distributed Unmanned-Aerial-Vehicular Networks for Small and Mini Drones," IEEE Veh. Technol. Mag. 12(3), 73-82 (2017).
57. H. L. Song and Y. C. Ko, "Beam Alignment for High-Speed UAV via Angle Prediction and Adaptive Beam Coverage," IEEE Trans. Veh. Technol. 70(10), 10185-10192 (2021).
58. Y. Cui, Q. Zhang, Z. Feng, *et al.*, "Specific Beamforming for Multi-UAV Networks: A Dual Identity-Based ISAC Approach," in *ICC 2023 - IEEE International Conference on Communications* (2023), pp. 4979-4985.